\documentclass{article}
\usepackage{spconf,amsmath,graphicx, bm}
\usepackage{amssymb}
\usepackage[bookmarks=false]{hyperref}
\usepackage{algorithm}
\usepackage{xcolor}
\usepackage{algpseudocode}
\usepackage{booktabs}
\usepackage{multirow}
\usepackage{caption} 
\usepackage{pifont}
\newcommand{\xmark}{\ding{55}}%
\title{Towards trustworthy phoneme boundary detection with autoregressive model and improved evaluation metric}
\name{Hyeongju Kim$^1$, Hyeong-Seok Choi$^{1,2}$}
\address{
$^1$Supertone, Inc.\\
$^2$Seoul National University}
\begin{document}
%\ninept
%
\maketitle
\begin{abstract}
Phoneme boundary detection has been studied due to its central role in various speech applications. In this work, we point out that this task needs to be addressed not only by algorithmic way, but also by evaluation metric. To this end, we first propose a state-of-the-art phoneme boundary detector that operates in an autoregressive manner, dubbed \textit{SuperSeg}. Experiments on the TIMIT and Buckeye corpora demonstrates that SuperSeg identifies phoneme boundaries with significant margin compared to existing models. Furthermore, we note that there is a limitation on the popular evaluation metric, R-value, and propose new evaluation metrics that prevent each boundary from contributing to evaluation multiple times. The proposed metrics reveal the weaknesses of non-autoregressive baselines and establishes a reliable criterion that suits for evaluating phoneme boundary detection.

\end{abstract}
\section{Introduction}
\label{sec:intro}

Phoneme-level segmentation of speech provides useful information for researching various speech applications such as speech recognition, speech synthesis, language identification, and singing voice synthesis. For instance, Liu~et~al.~\cite{liu2022pho} improved spoken language identification performance by aggregating latent variables between phoneme boundaries, and Lee~et~al.~\cite{lee2021voicemixer} showed promising results in voice style transfer by using variable-length segments that are basically similar to phoneme-level intervals.
% In light of the potential use of phoneme boundaries in many speech processing applications, it is necessary to develop an effective automatic system that can detect phoneme boundaries with high accuracy from a given speech source.
In light of the usability of phoneme boundaries in a broad range of speech processing, it is necessary to develop an effective automatic system that can detect phoneme boundaries with high accuracy from a given speech source.

Phoneme boundary detection is typically performed under two different settings depending on the presence of phoneme transcription. In a text-dependent scenario known as forced alignment, a pair of phonemes and an utterance are presented, and the start and end timestamps of each phoneme are estimated. This can be performed via an automatic speech recognition~(ASR) system based on a hidden Markov model~\cite{mcauliffe2017montreal}, attention alignments from a speech synthesis model~\cite{ren2019fastspeech}, or an explicit phoneme-to-audio aligner that is trained on supervisory signals obtained directly from phoneme boundaries~\cite{keshet2005phoneme,yuan2013automatic}. On the other hand, in a text-independent setting, the goal is to learn phoneme-level transitions in a speech source without knowledge of the corresponding phoneme sequence. Recent studies have proposed various methods for solving this problem in both supervised and unsupervised manners~\cite{kreuk2020phoneme,kamper2020towards, zhu2022phone}. 

We focus on developing a phoneme boundary detector for the \textit{text-independent} scenario under the \textit{supervised learning} setting. 
% The proposed model, namely \textbf{\textit{SuperSeg}}, employs an autoregressive architecture to leverage previous estimates of boundary detection. 
% The intuition is that additional information on whether the previous frames are classified as a boundary helps the model to resolve the uncertainty underlying in the frames near true boundaries thus leads to more reliable detecting results. 
First, we note that a naive classification approach is prone to produce multiple boundaries around a true boundary (i.e., over-segmentation). 
To tackle this, we propose \textbf{\textit{SuperSeg}} that employs an autoregressive architecture to leverage previous boundary estimates. SuperSeg prevents unwanted boundary repetitions by feeding additional information on whether the previous frames are classified as a boundary. Another issue is the volume of available data. In most cases, we have a smaller amount of data with phoneme boundary labels (e.g., TIMIT~\cite{garofolo1993timit}: 5.4 hours) compared to other speech datasets (e.g., %LibriTTS~\cite{zen2019libritts}: 585 hours,
MLS~\cite{Pratap2020MLSAL}: 60.5k hours) since phoneme boundary annotation is conducted at the frame-level with domain expertise. This lack of data often weakens the generalization performance for test data. To this end, we propose to adopt data augmentation techniques such as pitch/formant perturbations~\cite{choi2021neural} and masking blocks of frequency channels~\cite{park2019specaugment}. Furthermore, we propose new evaluation metrics for assessing phoneme-level segmentation. Previous metrics do not efficiently penalize duplicated estimates around true boundaries, thus overrating non-autoregressive baselines. Our proposed metrics provide a more trustworthy evaluation criterion by restricting the multiple contributions of each boundary to evaluation scores.

\section{Related work}
\label{sec:related_work}
There have been lots of studies for phoneme boundary detection. Frank~et~et.~\cite{franke2016phoneme} optimizes bidirectional LSTMs with a cross entropy function assigning more weight to phoneme boundary. Kreuk~et~al.~\cite{kreuk2020phoneme} employs learnable segmental features and identifies boundary candidates that minimizes the structured loss with dynamic programming. They also show that additional supervisory signals from phoneme labels yields improvements on boundary detection. Kamper~et~al.~\cite{kamper2020towards} utilizes the discrete code of pretrained vector quantized networks without using any phoneme boundary labels. Zhu~et~al.~\cite{zhu2022phone} adopts a contrastive learning scheme and a phoneme recognition task. Their model can perform forced alignment using a dynamic time warping algorithm as well as text-independent segmentation. Similarly, Kreuk~et~al.~\cite{kreuk2020self} uses the noise contrastive estimation and leverages a large amount of unlabeled audio data. Lin~et~al.~\cite{lin2022learning} employs a regularized attention mechanism on a pretrained acoustic encoder and performs text-dependent phoneme segmentation. 

For evaluation of phoneme boundary detection, R\"{a}s\"{a}nen et al.~\cite{rasanen2009improved} introduces R-value that is insensitive to random boundary insertion to penalize over-segmentation. To the best of our knowledge, R-value is by far the most reliable metric for assessing boundary segmentation.

\section{Proposed Method}
\label{sec:proposed_method}
\begin{figure}[t]
  \centering
  \includegraphics[width=0.8 \linewidth]{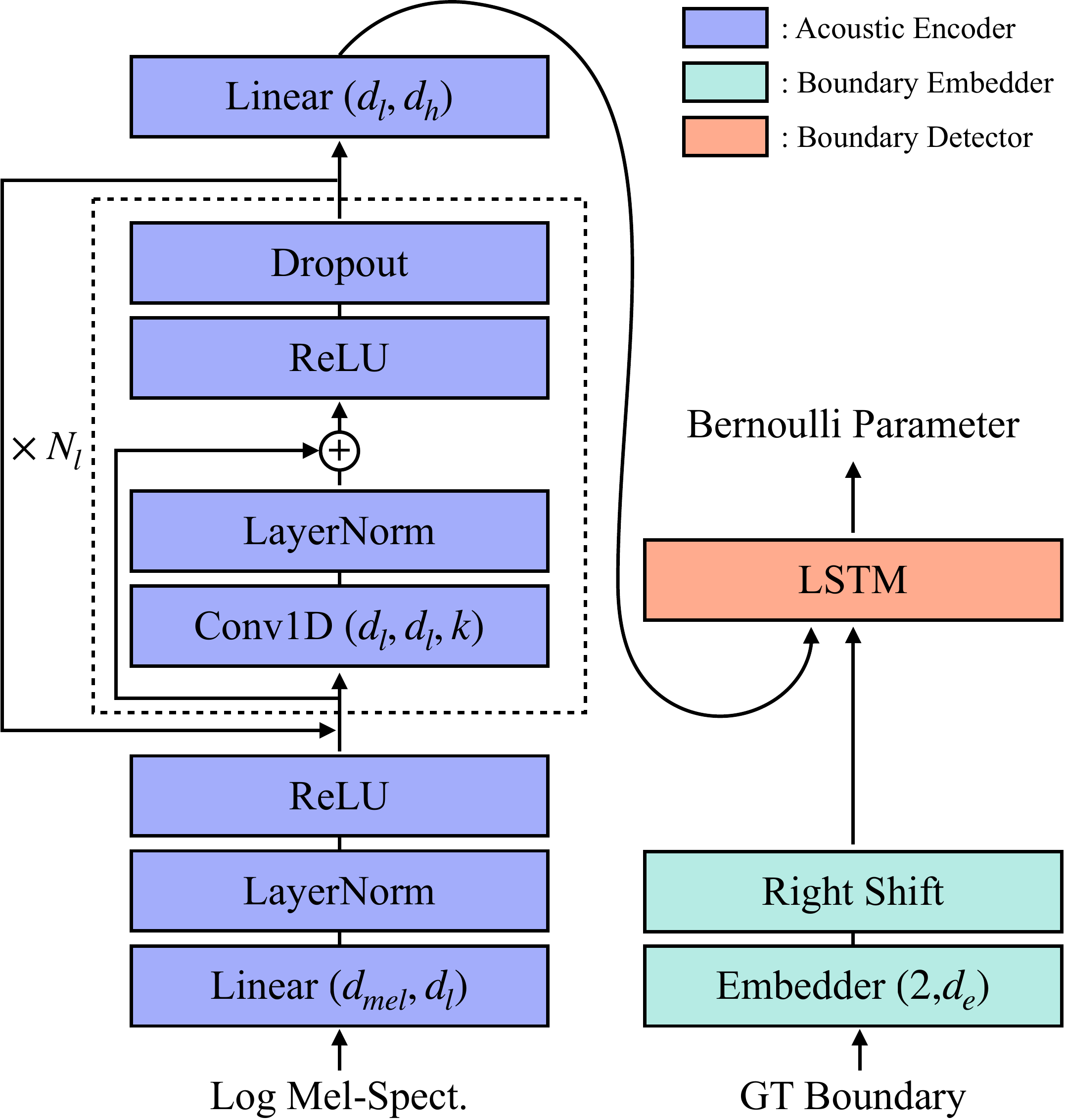}
  \caption{Detailed architecture of SuperSeg.}
  \label{fig:architecture}
\end{figure}

\begin{algorithm}[t]
\caption{Proposed evaluation algorithm}\label{alg:eval}
\begin{algorithmic}
\Require {$\gamma$ (tolerance), $A=\{a_i\}_{i=1,...,N_{A}}$ (sorted boundary list), $B=\{b_i\}_{i=1,...,N_{B}}$ (another sorted boundary list)}
\State{$N_{hit} \gets 0$}
\For{$a_{i} = a_{1}, a_{2}, ..., a_{N_{A}}$}
\For{$b_{j} \in B$}
\If{$|a_i - b_j| \le \gamma$}
\State{$N_{hit} \gets N_{hit} + 1$}
\State{\textcolor{red}{Remove $b_j$ from $B$}}
\State{\textbf{break}}
\EndIf
\EndFor
\EndFor
\State \Return{$\frac{N_{hit}}{N_{A}}$}
\end{algorithmic}
\end{algorithm}

\subsection{Architecture} SuperSeg consists of three main parts; acoustic encoder, boundary embedder, and boundary decoder. Acoustic encoder extracts latent features $\textbf{h}_1, ..., \textbf{h}_T \in \mathbb{R}^{d_h}$ from the mel-spectrogram of a given speech where $T$ denotes the total frame length. More specifically, acoustic encoder first transforms a log-scale mel-spectrogram of dimension $d_{mel}$ to $d_l$-dimensional variables $\textbf{m}_1, ..., \textbf{m}_T \in \mathbb{R}^{d_l}$ ($d_l \ge d_h$) and processes them using convolutional neural networks~(CNNs) consisting of ReLU activations, layer normalization~\cite{ba2016layer}, and dropout layers~\cite{srivastava2014dropout}. Boundary embedder maps binary values of boundary (1 if the current frame contains a phoneme boundary, 0 otherwise) to $d_{e}$-dimensional vectors $\textbf{e}_1, ..., \textbf{e}_T \in \mathbb{R}^{d_{e}}$. Boundary decoder employs a unidirectional LSTM~\cite{hochreiter1997long} that receives as inputs the $t$-th latent feature $\textbf{h}_t$ and the previous boundary embedding vector $\textbf{e}_{t-1}$, and outputs a Bernoulli parameter $p_t$ that estimates the probability that the current frame contains a phoneme boundary. Fig.~\ref{fig:architecture} shows the detailed architecture of SuperSeg.

\subsection{Training}
%We first annotate the binary labels for each frame of mel-spectrograms to indicate whether a phoneme boundary belongs or not.
% Then, 
SuperSeg is optimized to solve binary classification tasks using the binary cross entropy (BCE) loss. To satisfy the autoregressive nature, the boundary embedding vectors obtained from the ground-truth labels are shifted to the right by 1 time step, and passed to boundary decoder (i.e., teacher forcing). 
For better generalization, we adopt data augmentation algorithms such as pitch/formant perturbations~\cite{choi2021neural} and masking blocks of frequency channels~\cite{park2019specaugment}.

\subsection{Inference}
At test time, SuperSeg sequentially identifies phoneme-level boundaries on a frame-by-frame basis. 
To decide the label of the $t$-th frame, SuperSeg compares the model output $p_{t}$ with the threshold $\nu$ (1 if $p_{t} > \nu$ and 0 otherwise). The threshold $\nu$ is determined based on the evaluation metric (e.g., R-value~\cite{rasanen2009improved}) with the grid search using the validation set before evaluating on the test set.

\subsection{Evaluation metric}
\label{subsec:evaluation_metric}
% \begin{table*}[]
% \caption{\label{table:conventional}Comparison of phoneme boundary detection models on conventional evaluation metrics (\%).}
% \centering
% \begin{tabular}{l|cccc|cccc}
% \toprule
%                           & \multicolumn{4}{c|}{TIMIT}                                        & \multicolumn{4}{c}{Buckeye}                                       \\ \midrule
% \multicolumn{1}{c|}{Model} & Precision      & Recall         & F1-score       & R-value        & Precision      & Recall         & F1-score       & R-value        \\ \midrule
% Frank~et~al.~\cite{franke2016phoneme}               & 91.10          & 88.10          & 89.60          & 90.80          & 87.80          & 83.30          & 85.50          & 87.17          \\
% Kreuk~et~al.~\cite{kreuk2020phoneme}               & 94.03          & 90.46          & 92.22          & 92.79          & 85.40          & 89.12          & 87.23          & 88.76          \\
% Lin~et~al.~\cite{lin2022learning}                 & 93.42          & 95.96          & 94.67          & 95.18          & 88.49          & 90.33          & 89.40          & 90.90          \\
% SuperSeg (non-AR)          & \textbf{94.88} & \textbf{95.88} & \textbf{95.38} & \textbf{96.05} & \textbf{89.81} & \textbf{92.46} & \textbf{91.12} & \textbf{92.24} \\
% SuperSeg (AR)              & \textbf{95.63} & \textbf{94.77} & \textbf{95.20} & \textbf{95.82} & \textbf{89.92} & \textbf{89.94} & \textbf{89.93} & \textbf{91.40} \\ \bottomrule
% \end{tabular}
% \end{table*}
\begin{table*}[]
\caption{\label{table:conventional}Comparison of phoneme boundary detection models on conventional evaluation metrics (\%).}
\centering
\begin{tabular}{lc|cccc|cccc}
\toprule
        & & \multicolumn{4}{c|}{TIMIT}                                        & \multicolumn{4}{c}{Buckeye}    \\ \midrule
\multicolumn{1}{c|}{Model} & Use text & Precision      & Recall         & F1-score       & R-value        & Precision      & Recall         & F1-score       & R-value        \\ \midrule
\multicolumn{1}{l|}{Frank~et~al.~\cite{franke2016phoneme}}         &    \xmark  & 91.10          & 88.10          & 89.60          & 90.80          & 87.80          & 83.30          & 85.50          & 87.17          \\
\multicolumn{1}{l|}{Kreuk~et~al.~\cite{kreuk2020phoneme}}          & \xmark    & 94.03          & 90.46          & 92.22          & 92.79          & 85.40          & 89.12          & 87.23          & 88.76          \\
\multicolumn{1}{l|}{Lin~et~al.~\cite{lin2022learning}}             &  \checkmark  & 93.42          & 95.96          & 94.67          & 95.18          & 88.49          & 90.33          & 89.40          & 90.90          \\
\multicolumn{1}{l|}{SuperSeg (non-AR)} &   \xmark      & \textbf{94.88} & \textbf{95.88} & \textbf{95.38} & \textbf{96.05} & \textbf{89.81} & \textbf{92.46} & \textbf{91.12} & \textbf{92.24} \\
\multicolumn{1}{l|}{SuperSeg (AR)}   &  \xmark         & \textbf{95.63} & \textbf{94.77} & \textbf{95.20} & \textbf{95.82} & \textbf{89.92} & \textbf{89.94} & \textbf{89.93} & \textbf{91.40} \\ \bottomrule
\end{tabular}
\end{table*}

% Conventional metrics for rating the performance of phoneme boundary detection computes precision~(P), recall~(R), F1-score, and R-value from the ground-truth boundaries and the detected boundaries. 
\begin{figure}[t]
  \centering
  \includegraphics[width=1.0 \linewidth]{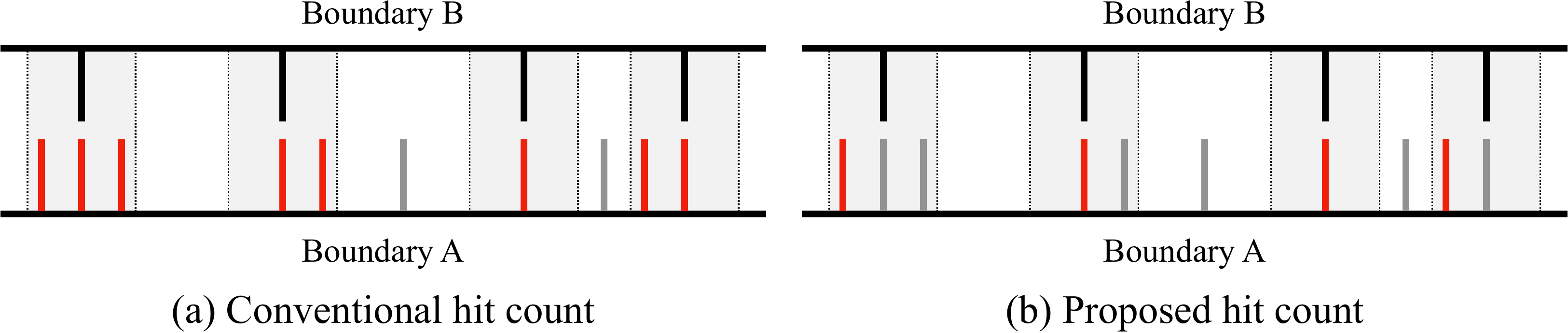}
  \caption{Comparison of hit counting methods. Hit and error boundaries are shown in red and gray, respectively.}
  \label{fig:hit_count}
\end{figure}
%Evaluation metrics such as F1-score and R-value are obtained from precision~(P) and recall~(R). 
In the conventional calculation of precision~(P) and recall~(R)~\cite{kreuk2020phoneme,zhu2022phone,kreuk2020self}, boundaries are usually evaluated not as a sequence but as individual elements, which overrates the repeated estimates around the true boundaries. We note that this often leads to unreliable scores of F1 and R-value since they are computed by using precision and recall.
%Although R-value~\cite{rasanen2009improved} is designed to penalize over-segmentation, it still has limitations in that it is basically measured by using precision and recall.
% In the calculations of these measurements, however, the boundaries are treated as individual elements, which overrates the repeated estimates around the true boundaries. %Also, our experiments reveals that R-value still has limitations although it is designed to penalize over-segmentation. 
%For example, suppose that the true label is [0, 0, 1, 0, 0] and the estimate is [0, 1, 1, 1, 0], and the tolerance level is one frame. In this case, the precision is 1 ($=\frac{3}{3}$) and the recall is 1 ($=\frac{1}{1}$), which does not reflect the actual accuracy. 
To tackle this issue, we propose to measure precision and recall by evaluating each boundary sequentially as shown in Alg.~\ref{alg:eval}. 
The proposed algorithm avoids multiple but redundant contributions, thus giving a low score for over-segmentation. We use the true and detected boundaries for $A$ and $B$ respectively to get the recall, and swap the order to calculate the precision. The F1-score and R-value are obtained from these precision and recall values in the same way as before. Fig.~\ref{fig:hit_count} demonstrates the conventional and proposed hit counting methods.
%Note that the proposed algorithm avoids the multiple contributions of each boundary to the evaluation score by excluding those already contributed. The results of the aforementioned example now becomes as follows; precision: $\frac{1}{3}$ and recall: 1.

\section{Experiments}
\label{sec:experiments}
\begin{figure}[t]
  \centering
  \includegraphics[width=0.95 \linewidth]{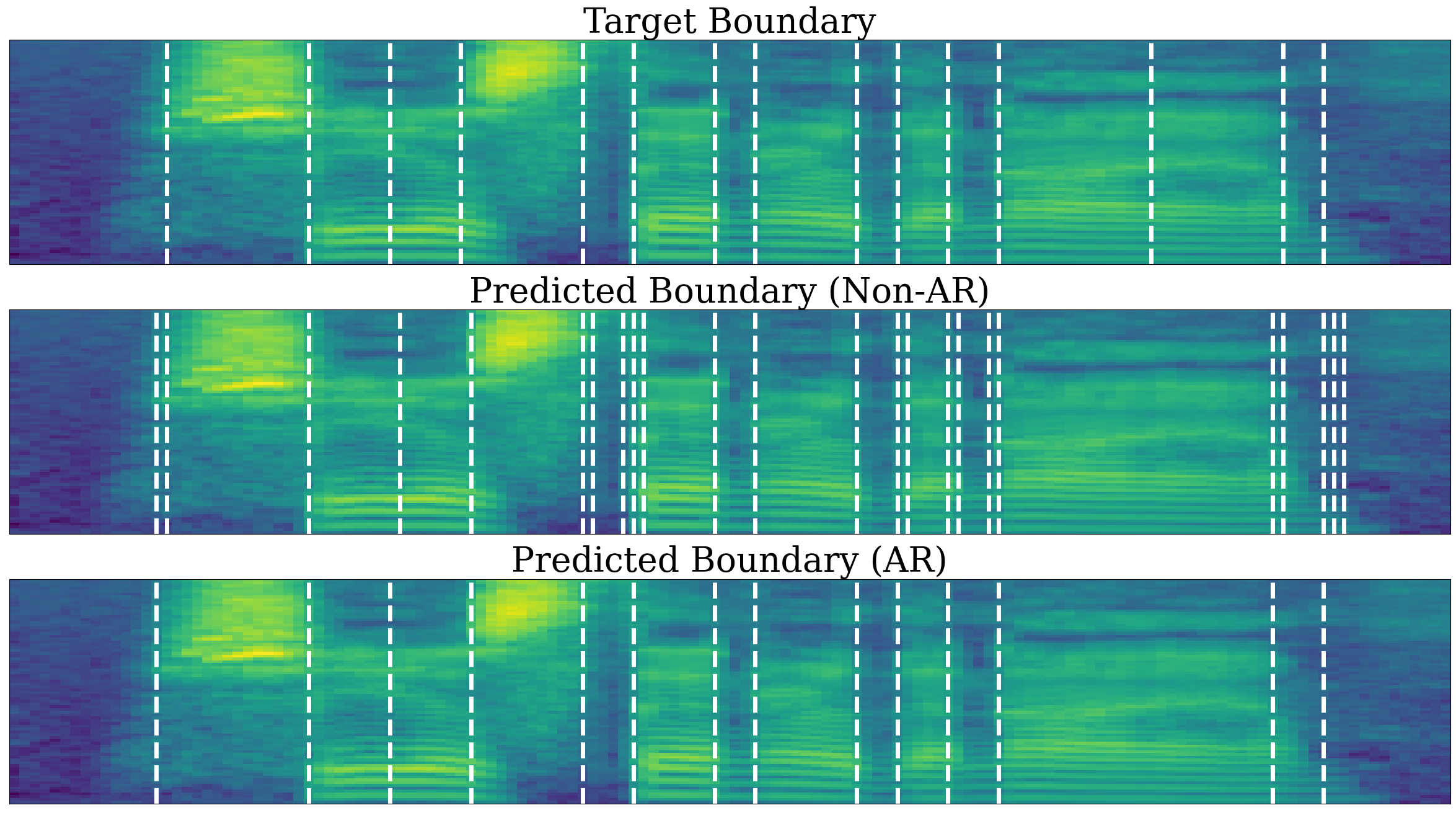}
  \caption{Example of boundary detection on TIMIT test data.}
  \label{fig:boundary}
\end{figure}

\subsection{Experimental setup}
We used as input a log-scale 80-channel mel-spectrogram ($d_{mel}=80$) that is computed on 40-millisecond windows with a stride of 10 milliseconds. The output dimension of the initial linear layer was set to 256 ($d_{l}=256$), and the dimension of the latent feature $\textbf{h}_{t}$ was set to 192 ($d_{h}=192$). We used 6 blocks  ($N_{l}=6$) consisting of a convolution network, layer normalization, ReLU activation, and a dropout layer with the zeroing probability of 0.4. The kernel size of convolution networks was set to 3 ($k=3$) and we used the dilation cycle of [1, 2, 4, 1, 2, 4]. The dimension of boundary embedding vector $\textbf{e}_{t}$ was set to 64 ($d_{e}=64$). Lastly, the unidirectional LSTM used hidden states with the size of 256 and transformed them into one-dimensional Bernoulli parameters.

We conducted experiments using two popular benchmarks for phoneme boundary detection; TIMIT~\cite{garofolo1993timit} and Buckeye~\cite{pitt2005buckeye} corpora. For the TIMIT dataset, we split the original training set into training and validation sets at a ratio of 9:1. The test split of the TIMIT dataset was used as is. For the Buckeye dataset, we constructed training, validation, and test sets by splitting the whole data at a ratio of 8:1:1 based on speaker identities following Kreuk~et~al.~\cite{kreuk2020self}.

We trained SuperSeg on the TIMIT and Buckeye datasets using the AdamW optimizer \cite{loshchilov2017decoupled} with a learning rate of $0.0005$ up to 1600 and 1000 epochs, respectively. The training was performed on a single RTX 3090 GPU with a batch size of 256. We employed data augmentation methods to reduce overfitting. For frequency masking, we uniformly sampled a mask size from $[0, 35)$ for every training data. For pitch/formant shift, we sampled a pitch multiplier from $(\frac{1}{1.2}, 1.2)$ and a formant multiplier from  $(\frac{1}{1.1}, 1.1)$.

\subsection{Results on conventional metrics}
\begin{table*}[]
\caption{\label{table:proposed}Comparison of phoneme boundary detection models on proposed evaluation metrics (\%). Model marked with $\dag$ denotes that evaluation is conducted on our  reimplementation. }
\centering
\begin{tabular}{l|cccc|cccc}
\toprule
                           & \multicolumn{4}{c|}{TIMIT}                                        & \multicolumn{4}{c}{Buckeye}                                       \\ \midrule
\multicolumn{1}{c|}{Model} & Precision      & Recall         & F1-score       & R-value        & Precision      & Recall         & F1-score       & R-value        \\ \midrule
Kreuk~et~al.~\cite{kreuk2020phoneme}$^\dag$         & 94.30          & 80.78          & 87.01          & 86.28          & 91.18          & 80.06          & 85.26          & 85.57          \\
SuperSeg (non-AR)          & 83.69          & 82.54          & 83.11          & 85.56          & 74.94          & 74.35          & 74.65          & 78.38          \\
SuperSeg (AR)              & \textbf{93.79} & \textbf{93.39} & \textbf{93.59} & \textbf{94.50} & \textbf{87.96} & \textbf{87.70} & \textbf{87.83} & \textbf{89.61} \\ \bottomrule
\end{tabular}
\end{table*}

\begin{table}[]
\caption{\label{table:ablation} Ablation study on data augmentation. A1 denotes frequency masking and A2 denotes pitch/formant perturbation.}
\centering
\begin{tabular}{c|cc|cccc}
\toprule
Corpus                   & A1       & A2       & P              & R              & F1             & R-val          \\ \midrule
\multirow{4}{*}{TIMIT}   &  \xmark        &   \xmark       & 92.20          & 92.18          & 92.19          & 93.33          \\
                         & \checkmark &   \xmark       & 93.66          & 93.34          & 93.50          & 94.44          \\
                         &    \xmark      & \checkmark & \textbf{94.16} & 93.33          & \textbf{93.75} & \textbf{94.59} \\
                         & \checkmark & \checkmark & 93.79          & \textbf{93.39} & 93.59          & 94.50          \\ \midrule
\multirow{4}{*}{Buckeye} &    \xmark      &   \xmark       & 86.08          & 85.67          & 85.87          & 87.93          \\
                         & \checkmark &   \xmark       & 86.18          & \textbf{88.25} & 87.21          & 89.01          \\
                         &   \xmark       & \checkmark & \textbf{88.65} & 86.90          & 87.77          & 89.44          \\
                         & \checkmark & \checkmark & 87.96          & 87.70          & \textbf{87.83} & \textbf{89.61} \\ \bottomrule
\end{tabular}
\end{table}

First, we compare the proposed method with other baselines using the conventional evaluation metrics. 
% Frank el al.~\cite{franke2016phoneme} and Kreuk~et~al.~\cite{kreuk2020phoneme} are working in the text-independent scenario, and Lin~et~al.~\cite{lin2022learning} performs forced alignment given a phoneme sequence. 
The methods proposed by Frank el al.~\cite{franke2016phoneme} and Kreuk~et~al.~\cite{kreuk2020phoneme} are working in the text-independent scenario, and Lin~et~al.~\cite{lin2022learning} performs forced alignment given a phoneme sequence. 
These baselines are all optimized under the supervised learning setting. We also trained and evaluated a non-autoregressive (non-AR) version of SuperSeg with boundary embedder excluded. In Table~\ref{table:conventional}, we report the results of precision (P), recall (R), F1-score (F1), and R-value (R-val) with a tolerance level of 20 milliseconds. The proposed model, SuperSeg (AR), achieves higher scores of F1-score and R-value than all the previous models in both TIMIT and Buckeye benchmarks. It's noteworthy that our \textit{text-independent} model trained solely on the TIMIT or Buckeye corpus outperforms the previous state-of-the-art \textit{text-dependent} model~\cite{lin2022learning} that leverages the pretrained wav2vec 2.0 model. The unexpected result is that, however, the non-AR model of SuperSeg shows better performance than the proposed AR model. In section~\ref{subsec:evaluation_metric}, we point out that the conventional evaluation criteria are vulnerable to the multiple but redundant contributions of adjacent boundaries for scores. To verify this, we investigated the boundaries predicted by the AR and non-AR SuperSeg models and the results are shown in Fig.~\ref{fig:boundary}. It can be observed that the non-AR model produces some repeated boundaries around the target boundaries to exploit the vulnerability of the evaluation metrics. On the other hand, the AR model detects phoneme boundaries appropriately without duplicated predictions.

\subsection{Results on proposed metrics}
Table~\ref{table:proposed} shows the results of several models using the proposed algorithm for computing precision and recall. First of all, the F1-score and R-value are all dropped compared to the scores in Table~\ref{table:conventional}. This is a preordained outcome since the proposed algorithm does not allow multiple commitments of each boundary to the evaluation scores. Secondly, the scores of the non-AR model are significantly reduced. This demonstrates that the non-AR model exploits the weaknesses of the conventional metrics and actually does not operate as desired. The other baseline~\cite{kreuk2020phoneme} also shows degraded performance on the proposed metrics. Thirdly, the proposed AR model shows decent performance comparable to the previous results with little decrease in the scores. Through this, we can verify that the proposed AR model produces trustworthy phoneme-level segments that fit our expectation.

\subsection{Ablation study}
To check the effect of frequency masking and pitch/formant perturbation, we additionally trained three SuperSeg models with different data augmentation settings. The experimental results are presented in Table~\ref{table:ablation}. When trained without data augmentation, the model performance becomes slightly degraded in both TIMIT and Buckeye corpora. Interestingly, when one of the data augmentations is used, the F1-score and R-value are comparable to or even better than the scores of the model trained with both techniques. This suggests that the data augmentation is useful for the training of SuperSeg to a certain extent.
 
 \section{Conclusion}
We introduced SuperSeg that builds on an autoregressive architecture to use previous model estimates as an additional input. %SuperSeg is a phoneme boundary detector designed for the text-independent scenario and is optimized under the supervised setting. 
To prevent the overfitting issue arising from the limited volume of the existing annotated datasets, we proposed to utilize data augmentation techniques such as frequency masking and pitch/formant perturbation. Furthermore, we proposed a new evaluation algorithm suitable for assessing phoneme boundary detection. Through the experiments, we showed SuperSeg achieves state-ot-the-art performance of phoneme boundary detection on both TIMIT and Buckeye corpora. We also demonstrated that the conventional metrics are vulnerable to the multiple contributions of a single boundary to a score and the proposed evaluation provides a reliable criterion by restricting these redundant commitments.

For future work, the text-dependent version of SuperSeg can be developed to find an alignment between a phoneme sequence and audio. Also, we expect that various speech applications will benefit from phoneme boundaries detected by SuperSeg. For instance, variable length voice conversion can be implemented by merging phoneme-level feature segments and expanding them back to different lengths based on the prosody of a target speaker.

% References should be produced using the bibtex program from suitable
% BiBTeX files (here: strings, refs, manuals). The IEEEbib.bst bibliography
% style file from IEEE produces unsorted bibliography list.
% -------------------------------------------------------------------------
\bibliographystyle{IEEEbib}
\bibliography{refs}

\end{document}